\documentclass[doublecol]{epl2}

\title{Piezomagnetism of FeSe single crystals}

\author{V. D. Fil\inst{1} \and D. V. Fil\inst{2} \and K. R. Zhekov\inst{1} \and T. N.
Gaydamak\inst{1} \and G. A. Zvyagina\inst{1} \and I. V.
Bilich\inst{1} \and D. A. Chareev\inst{3} \and A. N.
Vasiliev\inst{4,5}} \shortauthor{V. D. Fil \etal}

\institute{
  \inst{1} B. Verkin Institute for Low Temperature Physics
and Engineering National Academy of Sciences of Ukraine, 47 Lenin
Ave., Kharkov
61103, Ukraine\\
  \inst{2} Institute for Single Crystals, National Academy of Sciences of
Ukraine,  60 Lenin
Ave., Kharkov 61001, Ukraine\\
\inst{3}Institute of Experimental Mineralogy, Chernogolovka, Moscow Region, 142432, Russia\\
\inst{4}Low Temperature Physics and Superconductivity Department,
 Moscow State University, 119991 Moscow, Russia\\
 \inst{5}Theoretical Physics and Applied Mathematics Department, Ural Federal University
620002 Ekaterinburg, Russia

 } 

\abstract{ The acoustic-electric transformation in high-quality FeSe
single crystals is studied. In zero magnetic field we observe an
 {abnormally strong} electromagnetic radiation induced by a
transverse elastic wave. Usually a radiation of such intensity and
polarization is observed only in metals subjected to a high magnetic
field (the radiation is caused by the Hall current). We argue that
in FeSe in zero magnetic field it is caused by the piezomagnetic
effect  {which is most probably of dynamical origin}. We find that
the piezomagnetism survives under the transition from the normal to
superconducting state. In the superconducting state the
electromagnetic signal decreases with decreasing temperature that is
connected with the change in the London penetration depth.}

\begin{document}

\maketitle

An interplay between magnetic order and superconducting properties
of 11-type Fe-based superconductors remains  an open question. For
the most part, it is connected with the fact that the magnetic state
of such compounds is uncertain. For the first time, epitaxial films
of tetragonal FeSe were studied at room temperature and they behaved
as ferromagnets with saturation magnetization $\sim 500$
emu/cm$^3$\cite{1}. In \cite{1} it was also mentioned a
manifestation of the anomalous Hall effect (AHE) in FeSe films.
Unfortunately, any information on the superconducting properties of
these films was not obtained. The AHE was also observed in epitaxial
superconducting films FeSe$_{1-x}$Te$_x$ ($x = 0.5$) \cite{2}, but
the nature of the effect (ferromagnetic or antiferromagnetic) has
not been clarified. In the muon experiment \cite{3} it was found
that FeSe polycrystals acquire  {the static} magnetic order
(presumably antiferromagnetic) at enhanced ($>0.8$ GPa) pressure.
The nuclear magnetic resonance experiment on a polycrystalline
FeSe\cite{4} indicates  an increase in antiferromagnetic
fluctuations  close to the superconducting transition point. A weak
ferromagnetism with saturation magnetization $\sim 0.2$ emu/cm$^3$
was registered in FeSe single crystals \cite{5}. At the same time,
the neutron diffraction \cite{3} and Mossbauer experiments \cite{6}
did not reveal any magnetic order in FeSe polycrystals.

In all the papers cited above the samples were single-phase objects
(as confirmed by the X-ray analysis) and there is no reason to
attribute  magnetic ordering (or its tracks) to possible impurity
phases.  One should agree with the point of view of \cite{4} that
FeSe is very close to the transition into a magnetically ordered
state, and minor variations in the composition or internal stresses
may favor the transition (or, on the contrary, suppress it). In this
connection the study of perfect single crystals which composition
provides a high degree of homogeneity becomes of great importance.

In this letter we present the results of experiments on an
acoustic-electric transformation (AET) in high-quality  FeSe$_{0.963
\pm 0.005}$ single crystals. These results are rather intriguing. In
our opinion, they indicate the existence of the piezomagnetic (PZM)
effect in this crystal. It is commonly known \cite{7} that
piezomagnetism is only possible in a magnetically ordered state. We
argue that in our case it is the state with  {a dynamical} magnetic
order. In the superconducting state the AET signal decreases, but it
should be accounted for the change in the London penetration depth.
The piezomagnetic coefficient remains of the same value as in the
normal state.

\begin{figure}
\begin{center}
\includegraphics[width=8cm]{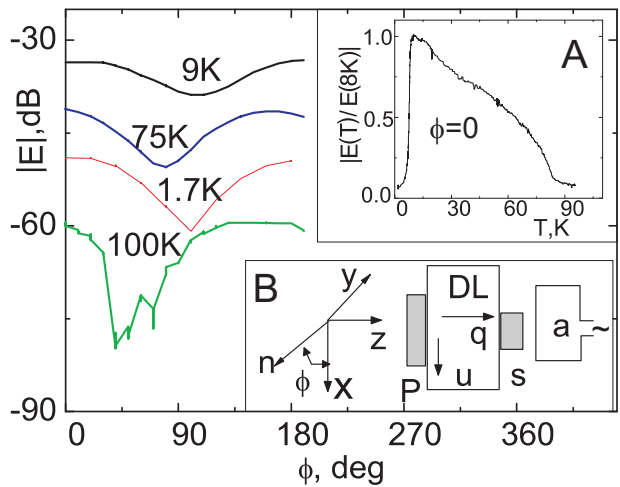}
\end{center}
\caption{The amplitude of the AET signals $|E|$ versus $\phi$, the
angle between the direction of the elastic displacement vector and
the normal to the plane of the receiving antenna. The inset A is the
temperature dependence of the amplitude at $\phi=0$. The inset B is
the experimental setup. Here P is the piezoelectric transducer, s is
the sample, DL is the delay line, a is the polarized antenna, and
${\bf n}$ is the normal to the antenna plane.} \label{f1}
\end{figure}

Let us briefly describe the  {theoretical grounds} for the AET
experiment. The experimental setup is shown in the inset  of Fig.
\ref{f1}. A more detailed information can be found in \cite{8,9,10}.
A transverse elastic wave with the wave vector ${\bf q}=(0,0,q)$ and
the displacement vector ${\bf u}= (u,0,0)$ enters into the sample
through a delay line. This wave can be considered as an alternating
and spatially modulated ion current. It produces an electromagnetic
( {EM}) field which forces free electrons to move to compensate the
initial disturbance, in accordance with the Le Chatelier principle.
For nonmagnetic samples the resultant current and the resultant
electrical component  of the  {EM field} are small by the ratio of
the electron mass to the ion mass (the inertial Stewart-Tolman
effect). They are aligned along the $x$-axis. In the
magnetic field ${\bf H} = (0, 0, H)$ 
the Lorentz force applied to free electrons results in the
appearance of the $y$-components of the current and the electrical
field (the Hall components). In view of lack of the compensating
forces the Hall components exceed considerable the inertial ones.
Due to continuity of the tangential components of  {the electric and
magnetic field} at the sample boundary the  {EM field} is radiated
from the surface. The radiation is registered by a polarized antenna
(a flat coil). The antenna registers the magnetic component of the
 {EM field} (its projection on the normal to the antenna
plane). Near the interface the  {EM field} is the plane wave and its
electrical component is equal the magnetic component. We will
discuss later only the electrical component $\mathbf{E}$. Its
projection on the antenna plane is referred to as the AET signal
$E$. It is the complex quantity and in our experiment we measure its
amplitude $|E|$ and phase $\mathrm{arg}(E)$ at different
orientations of the antenna.

The  {EM field} satisfies the Maxwell's equations that yield
\begin{equation}\label{1}
    \frac{d^2 \mathbf{E}}{d z^2}=\frac{4\pi
    i\omega}{c^2}\mathbf{j}+\frac{4\pi
    i\omega}{c}\nabla\times\mathbf{m},
\end{equation}
where ${\bf j}$ is the resultant current, and ${\bf m}$ is the
magnetic moment induced by the elastic wave.  The time dependence
$\propto\exp(i\omega t)$ is implied. For a nonmagnetic metal one can
use the  following local matter equation for the current
$${\bf j} =\hat{\sigma} ({\bf E}+{\bf W}),$$ where $\hat{\sigma}$
is the conductivity tensor the explicit form of which is well known
from the theory of galvanomagnetic phenomena\cite{11}
$$\hat{\sigma}=\sigma_0\left(
\begin{array}{cc}
1 & -\Omega \tau \\
\Omega \tau & 1 \\
\end{array}
\right)^{-1}$$($\sigma_0$ is the static conductivity, $\tau$ is the
relaxation time, and $\Omega$ is the cyclotron frequency), and ${\bf
W}=(U_{ST}, U_{ind},0)$ is the extraneous electromotive force
\cite{12} with $U_{ST} =(m/e)\omega^2 u$, the Stewart-Tolman field,
and $U_{ind}=\frac{i\omega}{c} B u$, the inductive field ($B$ is the
magnetic induction in the sample).

For  ${\bf m} = 0$ in the limit $\Omega\tau\ll 1$  the solution of
eq. (\ref{1})  has the form \cite{10}
\begin{equation}\label{2}
    \left(
      \begin{array}{c}
        E_x \\
        E_y \\
      \end{array}
    \right)=-\frac{k_0^2 }{q^2+k_0^2}\left(
    \begin{array}{c}
    U_{ST}- \Omega\tau\frac{q^2}{q^2+k_0^2} U_{ind} \\
                                           U_{ind}
                                         \end{array}
                                       \right),
\end{equation}
where $k_0^2=4\pi i\omega\sigma_0/c^2$ is the square of the
characteristic skin wave number. In deriving (\ref{2}) we neglect
the contribution of nonlocal effects (proportional to the electron
mean free pass, see \cite{8,10,12}) that are inessential for our
analysis.

For a superconductor in the Meissner state ($B=0$) the current
satisfies the London equation $i\omega {\bf
j}=(c^2/4\pi\lambda_L^2)({\bf E}+{\bf W})$, where $\lambda_L$ is the
London penetration depth. It yields
\begin{equation}\label{2-s}
E_x=-\frac{k_s^2 }{q^2+k_s^2} U_{ST},\quad E_y=0,
\end{equation}
where $k^2_s=\lambda_L^{-2}$.

Thus in zero magnetic field ($H = 0$) only the $E_x$ projection of
the electric field should be different from zero. Strictly speaking
this conclusion corresponds to the case where all the quantities
depend only on the $z$-coordinate (the one-dimensional case). The
actual situation is a three-dimensional one, but if the transverse
size of the sample exceeds the sound beam diameter the problem is
described satisfactorily as an effectively one-dimensional one.  In
the latter case the dependence of the amplitude of the AET signal on
the antenna orientation (the polarization diagram) demonstrates an
almost 100 percent modulation. In smaller samples the distortion of
the lines of the current  near the side surfaces results in the
appearance of the $E_y$ component  and in a reduction of the
modulation index, but the maximum keeps at the same orientation as
for large samples.

The measurements of the amplitude and phase of the AET signal can be
considered as a powerful tool for the study of  the vortex dynamics
in superconductors. It allows to obtain quantitative information on
the vortex viscosity, the vortex pinning strength \cite{9}, and the
Magnus force \cite{10}, and even to estimate the vortex
mass\cite{13}.

Our initial goal of the AET experiments with FeSe was the study of
the dynamics of the vortex matter in this compound. The single
crystals of FeSe were grown by the  technology,  {described in
\cite{14}. The samples under investigation were the (001) facet
platelets $\sim 1.5\times 1.5$ mm$^2$ in area. These samples were
used previously for the study of acoustic  characteristics
\cite{15}. The high quality of single crystals were confirmed by
indexing the X-ray diffraction patterns \cite{14} and by the
observation of the $\lambda$-anomaly in the heat capacity at the
superconducting transition \cite{15a}. The narrowness of the
superconducting transition ($\sim 0.5$K) seen from the width of the
longitudinal sound velocity jump \cite{15} also witnesses for the
sample quality.} The measurements have been done in the pulsed mode
at the frequency $\sim 55$ MHz. The pulse power and duration is
$10\div30$ W/cm$^2$ and 0.5 $\mu$s, correspondingly.  {The wave
vector was directed along the [001] axis of the sample}. The
diameter of the sound beam was $\sim 3$ mm. We use the equipment
described in \cite{16}. To minimize thermoelastic stresses that
emerge at the junction between the sample and the delay line a mylar
film is embedded into the junction.

Against all the expectations, in the absence of an external magnetic
field the recorded AET signal is in  two to three order larger than
one expected for the inertial field, and the maximum amplitude of
the AET signal corresponds to the electrical field ${\bf E}$
presumably polarized in the $y$ (not $x$) direction. The
polarization diagrams for several temperatures are shown in Fig.
\ref{f1}. As is expected for small samples the modulation index is
relatively small. The temperature dependence of the amplitude of AET
signal  in the $y$-direction ($\phi=0$) is shown in the inset A of
Fig. \ref{f1}. The signal appears below the temperature of
solidification of the silicone oil (GKZH-94) used as a bonding
material. Above  the superconducting transition the amplitude of the
signal grows monotonically under decrease in temperature. The
temperature dependence has a jump in the derivative at the point of
the tetra-ortho transformation in FeSe. In the superconducting state
the amplitude of the signal decreases.

\begin{figure}
\begin{center}
\includegraphics[width=8cm]{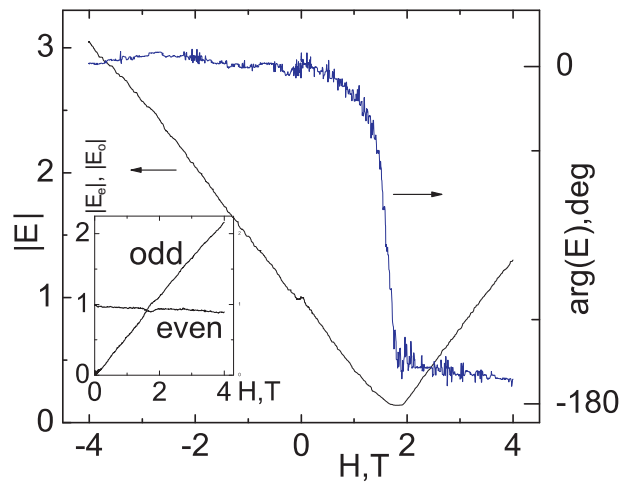}
\end{center}
\caption{The field dependence of the amplitude and phase of the AET
signal at $T=9$ K and $\phi=20^\circ$. The signal at $H_0=0$ is used
as the reference point for the phase and the scale for the
amplitude. The inset shows the amplitudes of the even and odd
components.} \label{f2}
\end{figure}

The magnetic field dependence demonstrates remarkable peculiarities.
For $\phi$ close to zero the dependence of the amplitude $|E|$ on
$H$ has a deep minimum. At $H$ that corresponds to the minimum of
$|E|$ the change in the phase of E is close to $\pi$. Such behavior
can be described in the vector diagram language.  The AET signal
behaves as a vector sum of two almost collinear components, an even
in the magnetic field and an odd one: $E_{e(o)}=[E(H)\pm E(-H)]/2$.
The amplitudes $|E_e|$ and $|E_o|$ are shown in the inset of Fig.
\ref{f2}. One can see that the odd component is linear in the
magnetic field and it is just the usual Hall component, while the
even one is practically independent of $H$. The amplitude of the
even component  is equal to the amplitude of the odd (Hall) one at
$B\approx 2$T. It allows us to estimate the value of the AET signal
at zero magnetic field. It is in the factor of $\Omega_{B= 2\rm{
T}}/\omega$ larger than one caused by the inertial Stewart-Tolman
force. For our frequency $\Omega_{B= 2\rm{ T}}/\omega\sim  10^3$.

Fig. \ref{f2} shows the most demonstrative example where two
components are in-phase or in antiphase. Under rotation of the
antenna the amplitudes and phases of two components vary in
different ways. Then, the position of the minimum is shifted and its
depth is changed. In principle, it is possible to choose the antenna
orientation for which the even and odd component are orthogonal to
each other. In that case the minimum disappears. The position of the
minimum depends on the quality of the surface treatment. In case of
crude treatment the minimum moves to the range of fields
inaccessible in our experiments ($>5.5$T). But in any case, the even
and odd components demonstrate a behavior similar to one shown in
the inset of Fig. \ref{f2}, only the relation between their modules
is changed. Let us also emphasize that we do not observe any
hysteresis in the field dependencies (shown in Fig. \ref{f2} as well
as obtained in other measurements), any nonlinear dependence of the
signal on the sound amplitude, and any step-like features (that
would be considered as a hallmark of spin-flop transitions).

Two effects may provide the Hall current in the absence of the
magnetic field. They are the AHE and  PZM effect and both of them
are realized in magnetically ordered media. As a rule the AHE is
observed in ferromagnetic conductors (see \cite{17} and references
therein), but, in principle, similar phenomenon is also possible in
antiferromagnets (AFM) \cite{18, 19}. The intensity of the AHE can
be characterized by the ratio of the off-diagonal component of the
conductivity tensor to the diagonal one:
$\eta=\sigma_{xy}/\sigma_{xx}$,  This ratio yields the fraction of
the transport current that branches off in a direction perpendicular
to it. There are many experiments on the AHE in ferromagnets
\cite{17}. Experimental data indicate that the maximum value of
$\eta$ does not exceed $10^{-2}$. Unfortunately, there is no
experimental statistics on the AHE in AFM. As far as we know the AHE
was observed only in hematite \cite{18}. For this compound the
branch off factor is $\eta\approx 2\times 10^{-4}$.

 The ferromagnetic AHE implies
that the sample has a spontaneous static magnetic moment $M$. Our
attempt to register the magnetic  {field} outside the samples was
unsuccessful (we used the fluxgate sensor  {with  sensitivity about
1 Oe} and examined one sample with dimension  $1.4 \times 1 \times
0.3$mm that corresponds to the demagnetizing factor $\approx 0.7$).
This means that even of the spontaneous moment exists, it is small
enough,
or it is concentrated in a thin surface layer that results in the
effective demagnetizing factor close to unity. 
In the latter case the thickness of the magnetic layer is of order
of the exchange force range $a_0\sim10^{-7}$ cm.

Let us check whether the AHE with physically relevant $\eta$
explains the observed anomaly. We write eq. (\ref{1}) for the $E_y$
component disregarding the contribution of the second term in the
right hand side of eq. (\ref{1}) and separating explicitly the
anomalous part of the current
\begin{equation}\label{3}
    \frac{d^2 E_y}{d z^2}=\frac{4\pi i \omega}{c^2}\left(\sigma_0
    E_y+j_{an}\right)=k_1^2 E_y,
\end{equation}
where $k_1^2\approx -q^2$ and $k_1^2\approx a_0^{-2}$ in case of
bulk and surface ferromagnetism, correspondingly, and
\begin{equation}\label{3-1}
   j_{an}=\eta(i\omega n e u )\frac{M}{M_{max}}
\end{equation}
is the anomalous current (here $M_{max}$  is the saturation magnetic
moment). The factor in the round brackets in eq. (\ref{3-1})
corresponds to the $x$-component of the electron current caused by
the sound beam (the analog of the transport current).  A possible
polydomain magnetic structure is accounted by the factor
$M/M_{max}$. For an AMF $k_1^2\approx -q^2$ and the expression for
the current (\ref{3-1}) contains the antiferromagnetic vector $L$
instead of $M$.

The quantity $\eta$ can be evaluated taking into account that at
$B=B_0\approx 2$ T the observed anomalous signal is of the same
value as the usual Hall signal (Fig \ref{f2}). From Eqs. (\ref {2}),
(\ref{3}) and (\ref{3-1}) we obtain
\begin{equation}\label{3-2}
    \eta=\left|\frac{k_1^2-k_0^2}{q^2+k_0^2}\right|\frac{\sigma_0B_0}{nec}\frac{M_{max}}{M}.
\end{equation}
 The resistivity of our
samples at $T=10$ K is $\rho\approx 40$ $\mu\Omega\cdot$cm
\cite{20a} that yields $|k_0^2|\approx 10^5$ cm$^{-2}$. The quantity
$q^2$ evaluated from the velocity of the $C_{44}$ mode
($s=1.38\times10^5$ cm/s \cite{15})  is $q^2 = 6.25\times 10^6$
cm$^{-2}$. The Hall measurements yield the density of carriers
$n\sim10^{20}\div 10^{21}$ cm$^{-3}$ \cite{1}, but since the
electron structure of FeSe corresponds to the compensated metal
\cite{20}, this quantity is most likely overestimated. A more
realistic estimate for $n$ can be obtained from the London
penetration depth: at $T=0$ $\lambda_L^2(0)=mc^2/4\pi n e^2$
\cite{11}. For the FeSe$_{0.94}$ polycrystal  the  {in-plane
magnetic penetration depth $\lambda_L(0)\approx 0.4\div 0.5$ $\mu$m}
was obtained in \cite{21}. From this value, assuming that $m$ is
close to the free electron mass, we evaluate $n\sim 10^{20}$
cm$^{-3}$. This yields the branch-off factor $\eta\sim 0.3$ (we put
the factor $M/M_{max}$ of order of unity). For our samples we obtain
even larger $\lambda_L(0)$ (see below) that corresponds to $n\sim
10^{19}$ cm$^{-3}$ and $\eta
> 1$. Assuming surface ferromagnetism we get $\eta\gg 1$. Such values of $\eta$ are unphysical and
we conclude that the anomalous AET behavior observed in our
experiment cannot be accounted for the AHE.

Let us now take into account the PZM effect. Neglecting $j_{an}$ and
considering the case $H=0$ we obtain from eq. (\ref{1})
\begin{equation}\label{4}
    \frac{d^2 E_y}{d z^2}=k^2 E_y- \frac{4\pi i \omega}{c}\frac{d m_x}{d
    z},
\end{equation}
where $k^2=k_0^2$ and $k^2=k_s^2$ for the normal and superconducting
state, correspondingly. In eq. (\ref{4}) $m_x$ is the $x$-component
of the magnetic moment induced by the elastic wave. It can be
presented as $m_x=\sum_i\Pi_{xixz} L_i C_{44} du/dz$ \cite{19},
where $C_{44}$ is the elastic constant, $L_i$ is the $i$-th
component of the vector of antiferromagnetism, and $\Pi_{xixz}$ are
the PZM coefficients (in \cite{7,22} the PZM modulus
 {$\Lambda_0$} defined as $\Lambda_0=\sum_i\Pi_{xixz} L_i$
was used). The general form for the matrices of the PZM coefficients
for different magnetic structures is given in \cite{19}.

Replacing  {$d^2/dz^2$ with $-q^2$}, we get
\begin{equation}\label{5}
    |E_y|=\frac{4\pi\omega}{c}\frac{q^2}{q^2+k^2} {\Lambda_0} C_{44} u.
\end{equation}
 {For FeSe $C_{44}\approx 10^{11}$ dyn/cm$^2$ \cite{15}.}
From the condition that the Hall signal at $B\approx 2$ T coincides
with the PZM signal  we obtain $ {\Lambda_0}= 2.4 \times 10^{-10}$
emu/(dyn$\cdot$cm). This is quite reasonable estimate that is  one
order less than the maximum known value for the PZM modulus
(measured in CoF$_2$ \cite{22}).

We would note that for the displacement vector polarized in the
$x$-direction the appearance of the $m_y$ component is not
principally prohibited, especially in polydomain samples. In the
latter case the PZM effect would be responsible for the appearance
of the $E_x$ component of the AET signal as well. In principle, it
may explain the temperature shift of the extremum in the
polarization diagrams.

 {The PZM effect considered is connected with that the
elastic deformations violate the strictly antiparallel configuration
of spins in AMF. In principle, another scenario that mimics the PZM
effect can be realized. Let us imagine that the sample has a nonzero
magnetic moment $\mathbf{M}$ which direction is bound to the [001]
axis by the anisotropy forces. The origin of the magnetic moment can
be the interstitial Fe or a magnetic phase grown at the surface
(while each time prior to the cooling the working surface of the
sample was cleaned by grinding with a fine abrasive powder, it does
not exclude the presence of another phase of the atomic thickness at
the surface). The elastic displacement $\bf{u}$ produces a tilt of
the [001] axis at the angle $\varphi=\mathrm{rot} \mathbf{u}/2$
\cite{7} with respect to the $z$ axis. Then the $m_x=M(du/dz)/2$
component appears. The measured value of $m_x$ is  $m_x=\Lambda_0
C_{44}(du/dz)$, where $\Lambda_0$ is given above. If the
interstitial Fe were responsible for the effect observed, the bulk
magnetization would be $M=2  \Lambda_0 C_{44} \approx 50$
emu/cm$^3$. Than, the magnetic field near the sample surface is
evaluated as $B_z=4\pi(1-b)M\approx 180$ G \cite{7} ($b=0.7$ is the
demagnetization factor). Such a field had to be registered easily by
the fluxgate sensor, but this was not the case. To provide the same
AET signal the magnetization of the magnetic phase at the surface
would be by the factor of $(qa_0)^{-1}>10^3$ larger (per unit
volume) than the bulk one. The latter possibility also looks
unrealistic.}

The explanation of our results in terms of the PZM effect
immediately raises the question on the type of the magnetic
structure in FeSe single crystals. In any case, it is obviously not
a  {static} AFM. It was shown in \cite{22} that for the usual AFM
the dependence of the PZM modulus on temperature reproduces the
temperature dependence of the sublattice magnetization. It follows
from (\ref{5}) that at $k_0^2\ll q^2$ the behavior of $|E_y|$ is
determined by the behavior of the PZM modulus. One can see from Fig.
\ref{f1} that in the normal state $|E_y|$ increases almost linearly
with the decrease of temperature, and it does not show any tendency
to the saturation. It is most likely that in case of FeSe we deal
with a  {dynamic} short-range magnetic order. This conclusion
correlates with an increase in AFM fluctuations in FeSe under
approaching Tc
 {\cite{4} and justifies the negative results of the
Mossbauer [6], neutron and $\mu$SR experiments [3]}. But we would
like to emphasize that   {for the observation of the PZM effect}
AFM fluctuations should have some preferable orientation of the
antiferromagnetic vector.

\begin{figure}
\begin{center}
\includegraphics[width=8cm]{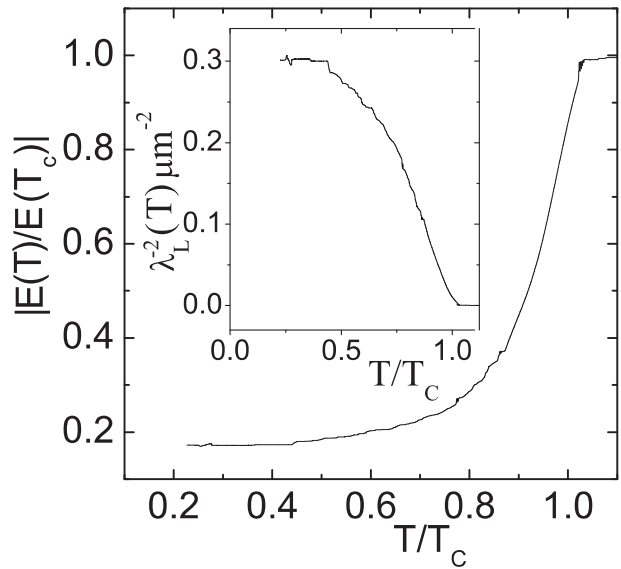}
\end{center}
\caption{The AET signal in the superconducting phase at $\phi=0$.
The inset shows the inverse square of the London penetration depth.}
\label{f3}
\end{figure}

The behavior of the AET signal in the superconducting state at $H =
0$ is shown in Fig. \ref{f3}. The signal  decreases below $T_c$, but
it does not disappear completely and saturates at the value easily
accessible for the measurements. Taking into account that for the
normal state $k^2=k_0^2\ll q^2$ and for the superconduction state
$k^2=k_s^2=\lambda_L^{-2}$ we get from eq. (\ref{5}) the relation
\begin{equation}\label{9}
 \frac{E_y^n(T_c)}{E_y^s(T)}=\left(1+\frac{1}{q^2
\lambda_L^2(T)}\right)f(T).
\end{equation}
Here the factor $f(T)$ accounts possible changes in the PZM modulus
in the superconducting state. Assuming $f(T)=1$ we obtain the
temperature dependence $\lambda_L^{-2}(T)$ shown in the inset of
Fig. \ref{f3}. The value $\lambda_L(0)$ obtained as the limit of
$\lambda_L(T)$ at $T\to 0$ is  $1.82 \pm 0.03$ $\mu$m. This quantity
can be also extracted from the slope of  $\lambda_L^{-2}(T)$ near
$T_c$ \cite{11}:
\begin{equation}\label{10}
    \lambda_L^{-2}(T)=\lambda_L^{-2}(0)\frac{2(T_c-T)}{T_c}
\end{equation}
that yields $\lambda_L(0)=(1.65 \pm 0.1)$ $\mu$m. These estimates
almost coincide to each other. This coincidence justifies our
assumption that the PZM modulus remains unchanged below $T_c$. In
other words, superconductivity and piezomagnetism co-exist
"peacefully". One can in principle think about some phase separation
scenario as a possible explanation of the behavior of the AET signal
in the superconducting state. Then neither expression (\ref{9}) nor
(\ref{10}) can be used for obtaining $\lambda_L(0)$, and
$\lambda_L(0)$ given by eq. (\ref{9}) can coincide with the one by
eq. (\ref{10}) only accidentally. Therefore, the phase separation
scenario looks doubtful.

We obtain $\lambda_L(0)$ that is in three times larger than one
found in \cite{21} for the FeSe$_{0.94}$ polycrystal, and our
estimate for  $\lambda_L(0)$ corresponds to the carrier density
$n\sim 10^{19}$ cm$^{-3}$. Perhaps the discrepancy with \cite{21}
can be accounted by for a strong dependence of the carrier density
on the structure of the sample and on its composition.

In summary, we have investigated  the acoustoelectric transformation
in high quality single crystals of FeSe. An  {abnormally strong}
electromagnetic radiation stimulated by the sound wave was detected
in zero magnetic field. Most likely the nature of the effect is
connected with the piezomagnetic properties of FeSe crystals. This
implies that FeSe has some kind of magnetic order  {most probably
the dynamical one}. The value of the piezomagnetic constant is
estimated. In the superconducting state the lowering of the AET
signal can be accounted completely for by the change in the London
penetration depth $\lambda_L(T)$. The latter means that the
piezomagnetic interaction remains unchanged. Our experiment yields a
rather large estimate for $\lambda_L(0)$, that is apparently due to
the low electron density.

\acknowledgments
This study was supported  in part by the grants of
Russian Foundation for Basic Research 12-02-90405, 13-02-00174, and
Russian Ministry of Science and Education 11.519.11.6012 and 8378.
The authors are grateful to N.F.Kharchenko, I.E.Chupis and
Yu.G.Pashkevich for the useful discussions and S.I.Bondarenko for
the assistance in the magnetic measurements.

\end{document}